\documentclass[onecolumn,10pt,showpacs,
amsmath,amssymb]{revtex4}
\usepackage{graphicx} 
\usepackage{hyperref} 
\usepackage{amssymb, amsmath}
\usepackage{epsf}

\def\la{\; \raise0.3ex\hbox{$<$\kern-0.75em\raise-1.1ex\hbox{$\sim$}}\;}
\def\ga{\;  \raise0.3ex\hbox{$>$\kern-0.75em\raise-1.1ex\hbox{$\sim$}}\;}

\def\pFn{p_{\raise-0.3ex\hbox{{\scriptsize F$\!$\raise-0.03ex\hbox{\rm n}}}}
}  % p_Fn

\def\pFa{p_{\raise-0.3ex\hbox{{\scriptsize F$\!$\raise-0.03ex\hbox{$i$}}}}
}  % p_{F i}

\def\pFas{p_{\raise-0.3ex\hbox{{\scriptsize F$\!$\raise-0.03ex\hbox{$k$}}}}
}  % p_{F i}

\def\pFb{p_{\raise-0.3ex\hbox{{\scriptsize F$\!$\raise-0.03ex\hbox{$\beta$}}}}
}  % p_{F \beta}

\def\vFa{v_{\raise-0.3ex\hbox{{\scriptsize F$\!$\raise-0.03ex\hbox{$i$}}}}
}  % v_{F i}
\def\pFp{p_{\raise-0.3ex\hbox{{\scriptsize F$\!$\raise-0.03ex\hbox{\rm p}}}}
}  % p_Fp
\def\pFe{p_{\raise-0.3ex\hbox{{\scriptsize F$\!$\raise-0.03ex\hbox{\rm e}}}}
}  % p_Fe
\def\pFmu{p_{\raise-0.3ex\hbox{{\scriptsize F$\!$\raise-0.03ex\hbox{\rm
$\mu$}}}} }  % p_Fe
\def\m@th{\mathsurround=0pt }
\def\eqalign#1{\null\,\vcenter{\openup1\jot \m@th
   \ialign{\strut$\displaystyle{##}$&$\displaystyle{{}##}$\hfil
   \crcr#1\crcr}}\,}

\newcommand{\vp}{\mbox{${\pmb p}$}}         % vector p (momentum)
\newcommand{\vps}{\mbox{${\vp '}$}}         % vector p (momentum)
\begin{document}

%%%%%%%%%%%%%%%%%%%%%%%%%%%%%%%%%%%%%%%%%%%%%%%%%%%%%%%%%%%%%%%%%%%%%%
\title{The relativistic entrainment matrix of a superfluid nucleon-hyperon 
mixture at zero temperature}
%%%%%%%%%%%%%%%%%%%%%%%%%%%%%%%%%%%%%%%%%%%%%%%%%%%%%%%%%%%%%%%%%%%%%%
%
\author{Mikhail E. Gusakov$^{1}$, Elena M. Kantor$^{1}$, Pawel Haensel$^{2}$}
\affiliation{
$^{1}$ Ioffe Physical Technical Institute,
Politekhnicheskaya 26, 194021 Saint-Petersburg, Russia
%e-mail: gusakov@astro.ioffe.ru
\\
$^{2}$ N. Copernicus Astronomical Center,
Bartycka 18, 00-716 Warsaw, Poland
}
\date{}
%

%\preprint{}

%%%%%%%%%%%%%%%%%%%%%%%%%%%%%%%%%%%%%%%%%%%%%%%%%%%%%%%%%%%%%%%%%%%%%%
\begin{abstract}
We calculate the relativistic entrainment matrix $Y_{ik}$
at zero temperature for nucleon-hyperon mixture 
composed of neutrons, protons, $\Lambda$- and $\Sigma^{-}$-hyperons,
as well as of electrons and muons.
This matrix is analogous to the entrainment matrix 
(also termed mass-density matrix or Andreev-Bashkin matrix) 
of non-relativistic theory.
It is an important ingredient for modelling the pulsations of 
massive neutron stars with superfluid nucleon-hyperon cores.
The calculation is done in the frame of the relativistic
Landau Fermi-liquid theory generalized 
to the case of superfluid mixtures;
the matrix $Y_{ik}$ is expressed through
the Landau parameters of nucleon-hyperon matter.
The results are illustrated with a particular example
of the $\sigma$-$\omega$-$\rho$ mean-field model 
with scalar self-interactions.
Using this model we calculate the matrix $Y_{ik}$
and the Landau parameters.
We also analyze stability of the ground state 
of nucleon-hyperon matter
with respect to small perturbations.
\end{abstract}
%%%%%%%%%%%%%%%%%%%%%%%%%%%%%%%%%%%%%%%%%%%%%%%%%%%%%%%%%%%%%%%%%%%%%%

\maketitle

%%%%%%%%%%%%%%%%%%%%%%%%%%%%%%%%%%%%%%%%%%%%%%%%%%%%%%%%%%%%%%%%%%%%%%%%%%%%%
%**************** Section 1 ******************************
\section{Introduction}
\label{1}
%%%%%%%%%%%%%%%%%%%%%%%%%%%%%%%%%%%%%%%%%%%%%%%%%%%%%%%%%%%%%

An analysis of electromagnetic \cite{sw06, sw07, lee08, tel08} 
and (in the future) gravitational \cite{ak01,andersson03,wkbs08}
radiation from pulsating neutron stars can shed light on 
the properties of superdense matter in their interiors.
The most interesting is the question about the composition 
of massive neutron-star cores 
(nucleons? hyperons? quarks? exotic matter?)
as well as about the properties of superfluid baryon matter
(the dependence of baryon critical temperatures on density, 
the type of pairing of various baryon species).

To interpret correctly the observational data
it is necessary to have realistic theoretical
models of pulsating neutron stars.
For that, one needs to formulate a hydrodynamics
which can be used to describe pulsations.
Clearly, the ordinary relativistic hydrodynamics
(see, e.g., \cite{ll87}), 
describing a liquid composed of identical particles, 
is not suitable for this purpose. 
The neutron-star cores are composed of a mixture
of various species of particles 
with baryons (nucleons and hyperons) 
that can be in superfluid state (\cite{ykgh01,yp04,lp04,pgw06,tnyt06}).
The hydrodynamics of superfluid mixtures 
strongly differs from the ordinary one because 
it allows the superfluid components to move independently 
of the normal (non-superfluid) liquid component without any 
dissipation of energy \cite{putterman74,khalatnikov89}.

This paper is devoted to a study 
of the relativistic entrainment matrix, 
which is an important quantity
in hydrodynamics of superfluid mixtures.
We mainly focus on the {\it nucleon-hyperon} matter 
in the core of massive neutron stars.
Notice that, until now 
only the superfluid hydrodynamics of nucleon matter,
composed of neutrons ($n$), protons ($p$), 
and electrons ($e$) 
with a possible admixture of muons ($\mu$),
has been considered in astrophysical literature.
Let us discuss the results of previous works in more detail.

Assume that the neutrons and protons are in superfluid state.
%In this case a motion with three independent velocities 
%is possible in nucleon matter.
In this case three independent velocities can exist in nucleon matter.
Two of them are the velocities 
${\pmb V}_{{\rm s}n}$ and ${\pmb V}_{{\rm s}p}$
of neutron and proton superfluid components, respectively.
The other is the velocity ${\pmb V}_{\rm qp}$ 
of `normal' (non-superfluid) neutrons and protons, 
as well as electrons and muons
(it is assumed that, due to collisions,
it is the same for all `normal' particles).
The physical meaning of the phenomenological superfluid velocities
${\pmb V}_{{\rm s}n}$ and ${\pmb V}_{{\rm s}p}$
can be understood on the basis of microphysics
(see, e.g., \cite{gh05} and Sec.\ IIB).
It turns out that the velocity ${\pmb V}_{{\rm s}i}$
is related to a Cooper pair momentum $2 \, {\pmb Q}_i$ 
of nucleon species $i=n,p$ by the equality
\begin{equation}
{\pmb V}_{{\rm s}i} = \frac{{\pmb Q}_i}{m_i},
\label{VsQ}
\end{equation}
where $m_i$ is the mass of a free paricle species $i$.

The non-relativistic expressions for the mass current density
of neutrons ${\pmb J}_n$ and protons ${\pmb J}_p$ have the form
(see, e.g., \cite{bjk96,gh05,ch06})
\begin{eqnarray}
{\pmb J}_n &=& (\rho_n - \rho_{nn} - \rho_{np}) {\pmb V}_{\rm qp}
+ \rho_{nn} {\pmb V}_{{\rm s}n} + \rho_{np} {\pmb V}_{{\rm s}p},
\label{Jn} \\
{\pmb J}_p &=& (\rho_p - \rho_{pp} - \rho_{pn}) {\pmb V}_{\rm qp}
+ \rho_{pp} {\pmb V}_{{\rm s}p} + \rho_{pn} {\pmb V}_{{\rm s}n}.
\label{Jp}
\end{eqnarray}
Here $\rho_n$ and $\rho_p$ 
are the neutron and proton density, respectively;
$\rho_{ik}=\rho_{ki}$ is the symmetric $2 \times 2$ 
entrainment matrix, also termed Andreev-Bashkin 
or mass-density matrix ($i$, $k=n$, $p$). 
It follows from Eqs.\ (\ref{Jn}) and (\ref{Jp})
that superfluid motion of, for example, neutrons,
contributes not only to ${\pmb J}_n$ but also to ${\pmb J}_p$ 
(and the same for protons).
For the first time this effect was predicted, 
as applied to superfluid solutions of $^3$He in $^4$He,
by Andreev and Bashkin \cite{ab75}.
The prefactors in front of ${\pmb V}_{\rm qp}$ 
in Eqs.\ (\ref{Jn}) and (\ref{Jp}) can be interpreted
as the densities of `normal' neutrons and protons, 
respectively.
Since at zero temperature ($T=0$) all particles are paired,
these densities vanish and we have \cite{bjk96,gh05,ch06}
\begin{eqnarray}
\rho_n &=& \rho_{nn}+\rho_{np},
\label{111}\\
\rho_p &=& \rho_{pp}+\rho_{pn}.
\label{sumrule_nonrel}
\end{eqnarray}
More strictly these conditions can be obtained 
from the requirement of Galilean invariance of 
the equations of superfluid hydrodynamics at $T=0$ \cite{bjk96,ch06}.

The matrix $\rho_{ik}$ was calculated for the case of $T=0$
in Refs.\ \cite{bjk96,ch06} and for arbitrary temperatures 
in Ref.\ \cite{gh05}. 
In both cases, the authors used the non-relativistic 
Fermi-liquid theory of Landau.
Though neutrons and especially protons in the cores of low-mass 
neutron stars can be (with a reasonable accuracy) considered 
as non-relativistic, more self-consistent 
(and necessary in the case of massive neutron stars) is the approach, 
in which nucleons are treated in the frame of
relativistic theory.
Following Refs.\ \cite{son01,ga06,gusakov07,gk08}, 
the relativistic analogue of Eqs.\ (\ref{Jn}) and (\ref{Jp})
can be presented in the form
\begin{equation}
{\pmb j}_i = \left( n_i-\sum_k \, \mu_k \, Y_{ik} \right) {\pmb u} 
+ c^2 \sum_k \, Y_{ik} \, {\pmb Q}_k.
\label{jnn}
\end{equation}
Here ${\pmb j}_i$ 
is the particle current density ($i=n, p$);
$c$ is the speed of light;
${\pmb u}$ is the spatial component 
of the four-velocity $u^{\mu}$, 
normalized by the condition $u^{\mu} u_{\mu} = -c^2$, 
and describing the motion of normal part of liquid;
$n_i$ and $\mu_i$ are, respectively, the number density and 
relativistic chemical potential of particle species $i$,
measured in the frame where $u^{\mu}=(c,0,0,0)$.
Finally, the symmetric matrix $Y_{ik}$ 
is the relativistic analogue
of the entrainment matrix $\rho_{ik}$.
In the non-relativistic limit Eq.\ (\ref{jnn}) is equivalent
to Eqs.\ (\ref{Jn}) and (\ref{Jp}) under conditions that
\begin{equation}
{\pmb u}={\pmb V}_{\rm qp}, \quad \quad
\rho_{ik} = m_i m_k \, c^2 \, Y_{ik},
\label{nonrel_limit}
\end{equation}
%
%where $c$ is the speed of light.
The prefactor in front of ${\pmb u}$ in Eq.\ (\ref{jnn})
can be interpreted 
(by analogy with the non-relativistic case)
as the number density of `normal' (non-superfluid)
particle species $i$.
At zero temperature this number density vanishes.
This imposes a condition on the matrix $Y_{ik}$ \cite{ga06}
\begin{equation}
\sum_k \mu_k \, Y_{ik} = n_i.
\label{sumrule}
\end{equation}
Taking into account this condition, Eq.\ (\ref{jnn})
can be rewritten (at $T=0$) in the form
\begin{equation}
{\pmb j}_i = c^2 \sum_k \, Y_{ik} \, {\pmb Q}_k.
\label{entrmatr}
\end{equation}
The matrix $Y_{ik}$ for matter composed of
neutrons and protons was calculated at $T=0$
in Ref.\ \cite{cj03} 
(the authors of Ref.\ \cite{cj03} 
used a somewhat different formalism, 
see, e.g., the review \cite{ac07} and references therein).
The calculation was done in the frame of 
relativistic $\sigma$-$\omega$ mean-field model.

This paper is a natural continuation of the research described above.
Our aim is to calculate the relativistic entrainment matrix $Y_{ik}$
at zero temperature for matter composed 
not only of nucleons, electrons, and muons, but also of hyperons.
We consider only two types of hyperons, 
namely $\Lambda$- and $\Sigma^{-}$-hyperons 
(to be denoted by $\Lambda$ and $\Sigma$, respectively).
In most of the calculations, presented in the literature, 
they appear first in the neutron-star matter 
with the increasing density (see, however, \cite{fg07}). 
We wish to emphasize, 
that analytical results, 
obtained in this paper, can be (in principle) 
applied to any number of superfluid baryon species.

At this point it is convenient to 
make a few remarks
%reservations 
%present a short discussion
concerning the hyperon interactions and superfluidity.
%Let us discuss in some more detail the interactions 
%and superfluidity of hyperons.
First, recent experiments indicate 
that the interaction of $\Sigma^-$-hyperons with nucleons
is repulsive (see, e.g., Ref.\ \cite{fg07}).
For a sufficiently strong repulsion it is possible that 
$\Sigma^-$-hyperons 
%appear at higher densities 
%than it was thought before
%or even do 
may not appear in the neutron stars at all.
In this case they can be (in some models) 
`replaced' by $\Xi^-$-hyperons 
(see the discussion of Fig.\ 11 in Ref.\ \cite{fg07}).
However, in the case of not very strong repulsion, 
$\Sigma^-$-hyperons can appear 
rather close to a density threshold for $\Lambda$-hyperons
(see, e.g., Refs.\ \cite{bg04, dsw08}.
%The repulsion can also lower the critical temperature of $\Sigma^-$-hyperons.
The repulsion can also shift the critical temperature of $\Sigma^-$-hyperons
from $(5 \times 10^{10}-5 \times 10^{11})$ K 
(see, e.g., \cite{bb98, vt04, tnyt06}) to a lower value.
%This problem is not sufficiently studied in the literature.

Second, as suggested by the Nagara event \cite{takahashi01}, 
$\Lambda\Lambda$ attraction can be weaker than it was assumed before.
This may result in substantial decreasing 
of critical temperature of $\Lambda$-hyperons 
(from $(10^9-10^{10})$ K to a temperature much less than $10^8$ K, 
see Ref.\ \cite{tnyt06}). 
%and even make $\Lambda$-hyperons nonsuperfluid. 
However, it is too early to draw a final conclusion 
(see, e.g., the criticism of this result on p. 23 
of Ref.\ \cite{tnyt06}).
The real interaction may be stronger 
than that deduced from the event 
(and, of course, new experimental evidences are necessary).

Even if we assume the weak attraction of $\Lambda$-hyperons 
and the repulsion between $\Sigma^-$-hyperons and nucleons, 
it is possible that the inclusion of in-medium effects 
or three-body forces may significantly influence 
(increase or further decrease)
the hyperon critical temperatures.
Morever, the exotic pairing 
of $\Lambda$ and $\Sigma^-$-hyperons 
may take place \cite{tt08}.
Taking into account the above discussion, 
it is reasonable to treat hyperon critical temperatures 
as free parameters.
Since in this paper we consider the case of zero temperature,
below we assume that all baryon species are superfluid.

The phenomenological equations 
(\ref{VsQ}) and (\ref{jnn})--(\ref{entrmatr}),
which are discussed in this section 
in the context of nucleon matter,
remain unchanged for nucleon-hyperon matter.
The only difference is that now the indices $i$ and $k$
run over $i, k = n$, $p$, $\Lambda$, $\Sigma$ 
(see also Ref.\ \cite{gk08}).
Thus, now $Y_{ik}$ is a $4 \times 4$ matrix.

The paper is organized as follows.
In Sec.\ II the {\it relativistic} Landau Fermi-liquid theory
\cite{bc76} is generalized to the case of superfluid mixtures.
In the frame of this theory 
we calculate the matrix $Y_{ik}$ 
and express it through the Landau parameters 
$f_1^{ik}$ of nucleon-hyperon matter.
In Sec.\ III the general results of Sec.\ II are illustrated
with a particular example of the $\sigma$-$\omega$-$\rho$
mean-field model with scalar self-interactions \cite{glendenning85}. 
Namely, we 
$(i)$ calculate the matrix $Y_{ik}$;
$(ii)$ determine {\it all} (spin-averaged) Landau parameters,
corresponding to this model;
$(iii)$ analyze the stability of the ground state
of nucleon-hyperon matter with respect to small perturbations.
Section IV contains a summary of our results.

%%%%%%%%%%%%%%%%%%%%%%%%%%%%%%%%%%%%%%%%%%%%%%%%%%%%%%%%%%%%%%%%%%%%%%%
\section{The relativistic entrainment matrix at zero temperature 
from the Landau Fermi-liquid theory}
\label{2}
%%%%%%%%%%%%%%%%%%%%%%%%%%%%%%%%%%%%%%%%%%%%%%%%%%%%%%%%%%%%%%%%%%%%%%%

%%%%%%%%%%%%%%%%%%%%%%%%%%%%%%%%%%%%%%%%%%%%%%%%%%%%%%%%%%%%%%%%%%%%%%%
\subsection{Relativistic Landau theory for mixtures of Fermi liquids}
\label{2a}
%%%%%%%%%%%%%%%%%%%%%%%%%%%%%%%%%%%%%%%%%%%%%%%%%%%%%%%%%%%%%%%%%%%%%%%

In this section we briefly discuss how to generalize
the Landau Fermi-liquid theory to the case of nucleon-hyperon 
mixture composed of neutrons, protons, $\Lambda$-, 
and $\Sigma^{-}$-hyperons.
The original non-relativistic Landau Fermi-liquid theory 
(e.g., \cite{lp80, bp91})
was extended to the case of mixtures of protons and neutrons
in the paper by Sj$\ddot{ {\rm o}}$berg \cite{sjoberg76} 
(see also \cite{bjk96}).
The relativistic generalization of the Landau Fermi-liquid 
theory was given by Baym and Chin \cite{bc76} 
(they considered a Fermi-liquid composed of identical particles).

The generalization of the Landau theory to the case 
of relativistic mixtures composed of 
more than one component can be made
in the same way as in Refs.\ \cite{sjoberg76,bc76}.
Thus, we only briefly describe the main formulae of the theory 
which will be used subsequently.
Notice, that the results obtained below in Section II
can be applied to a Fermi-liquid composed 
of any number of baryon species (not necessarily four).
In this case the particle indices in equations
should run over all baryon species.

Unless is stated otherwise, throughout 
the rest of the paper we use the system of units
in which the Planck constant $\hbar$, 
the Boltzmann constant $k_{\rm B}$, 
and the speed of light $c$ equal unity, 
$\hbar=k_{\rm B}=c=1$.
We also imply that the subscripts $i$ and $k$ refer to baryons.

As it is demonstrated in Ref.\ \cite{bc76}, generally,
the structure of the relativistic Landau theory 
of Fermi liquids is the same
as that of the non-relativistic theory.
Only those results of both theories are different
which are obtained using Lorentz (Galilean) 
transformation properties of various quantities 
(in particular, the energy and momentum).
For instance, we will show that 
the relativistic expression 
for the effective mass of particle species $i$
differs from its non-relativistic analogue.

Let us consider a system in the ground state 
with the energy $E_{0}$ at temperature $T=0$.
The distribution 
function of quasiparticle species $i$ 
is then a Fermi sphere,
\begin{equation}
n_{i0}({\pmb p}) = \theta(\pFa - p),
\label{Fermisphere}
\end{equation}
where ${\pmb p}$ is the quasiparticle momentum;
$\theta(x)$ is the step function: $\theta(x) =1$, 
if $x>0$ and $0$ otherwise.
A small deviation $\delta n_i({\pmb p})$ 
of the distribution function from $n_{i0}({\pmb p})$
changes the system energy by
\begin{equation}
E-E_{0} =  
\sum_{{\pmb p} s i} \varepsilon_{i0}({\pmb p}) 
\, \delta n_i({\pmb p})
+ \frac{1}{2} \sum_{{\pmb p} {\pmb p}' \, ss' \, ik} 
f^{ik}({\pmb p}, {\pmb p}') \, 
\delta n_i({\pmb p}) \delta n_k({\pmb p}').
\label{energy2}
\end{equation}
Addition of one more quasiparticle of a species $i$, 
with the momentum ${\pmb p}$, to the system, 
increases the total energy $E$ by the energy
$\varepsilon_i({\pmb p})$ of the quasiparticle. 
From Eq.\ (\ref{energy2}) it follows that
\begin{equation}
\varepsilon_{i}({\pmb p}) =\varepsilon_{i0}({\pmb p})
+ \sum_{{\pmb p}' s' k} f^{ik}({\pmb p}, {\pmb p}') \, 
\delta n_k({\pmb p}'). 
\label{energy_quasi}
\end{equation}
%
%
%\begin{equation}
%\varepsilon_{i}({\pmb p}) =\varepsilon_{i0}({\pmb p})
%+ \sum_{{\pmb p}' s' k} f^{ik}({\pmb p}, {\pmb p}') \, 
%\delta n_k({\pmb p}'). 
%\label{energy_quasi}
%\end{equation}
%
In Eqs. (\ref{energy2}) and (\ref{energy_quasi}) 
$\vp$ and $\vps$ are the particle
momenta; $s$ and $s'$ are the spin indices;
$i, k = n$, $p$, $\Lambda$, and $\Sigma$ 
are the baryon species indices.
Furthermore, $\varepsilon_{i0}({\pmb p})$ is the energy
of a quasiparticle of a species $i$, corresponding to 
the distribution function $n_{i0}({\pmb p})$.
It can be expanded into a series near the Fermi surface
in powers of the quantity $p-\pFa$ and presented in the linear form
\begin{equation}
\varepsilon_{i0}({\pmb p}) \approx \mu_i+ \vFa (p-\pFa),
\label{energy0}
\end{equation}
where $\pFa$ is the Fermi momentum of (quasi)particle species $i$;
$\mu_i= \varepsilon_{i0}(\pFa)$ is the relativistic chemical potential
or, equivalently, the Fermi energy of quasiparticle species $i$;
${\pmb v}_{{\rm F}i} 
= [\partial \varepsilon_{i0}({\pmb p})/\partial{\pmb p} ]_{p=\pFa}$
is the velocity of quasiparticles on the Fermi surface.
It can also be expressed as $\vFa \equiv \pFa/m^{\ast}_i$,
where $m^{\ast}_i$ is the effective mass of quasiparticle species $i$.
%
%We restrict ourselves to a spin-unpolarized nucleon matter.
%This allows us to simplify the notations.
%For instance, we will drop spin
%indices, whenever possible.
Finally, the function $f^{i k}({\pmb p}, {\pmb p}')$
in Eq.\ (\ref{energy_quasi})
is the spin-averaged
Landau quasiparticle interaction
(here and below we disregard the spin-dependence of this interaction
since it does not affect our results).
In the vicinity of the Fermi surface the arguments of
the function $f^{i k}({\pmb p}, {\pmb p}')$
can be approximately put equal to $p \approx \pFa$ 
and $p' \approx \pFas$, 
while the function itself can be 
expanded into Legendre polynomials $P_l({\rm cos \, \theta})$,
\begin{equation}
f^{i k}({\pmb p}, {\pmb p}') = \sum_l f^{i k}_l \, P_l(\cos \theta),
\label{fik}
\end{equation}
where $\theta$ is the angle between ${\pmb p}$ and ${\pmb p}'$
and $f^{i k}_l$ are the (symmetric) Landau parameters,
$f_l^{ik}=f_l^{ki}$.

As in the non-relativistic case, the effective mass $m_i^{\ast}$ 
in the relativistic theory can be expressed in terms of 
the Landau parameters $f^{i k}_1$.
To find this relation let us consider, following Ref.\ \cite{bc76}, 
two frames  $K$ and $\overline{K}$, and assume that the frame 
$\overline{K}$ moves with the velocity ${\pmb V}$ 
with respect to $K$.
Below in this section the quantities marked with an overline
will be referred to the frame $\overline{K}$, while those 
without the overline -- to the frame $K$.
The total energy of nucleon-hyperon mixture $E$ ($\overline{E}$) 
and its momentum ${\pmb P}$ ($\overline{\pmb P}$) 
are related by the Lorentz transformation
\begin{eqnarray}
E &=&(\overline{E}+\overline{{\pmb P}}{\pmb V}) \, \gamma,
\label{EEE}\\
{\pmb P} &=& \overline{{\pmb P}} - {\pmb e_{\pmb V}} 
\left({\pmb e_{\pmb V}} \overline{{\pmb P}} 
\right)\, (1- \gamma) +\overline{E} {\pmb V}\gamma.
\label{PPP}
\end{eqnarray}
In Eqs.\ (\ref{EEE}) and (\ref{PPP})
$\gamma=(1-V^2)^{-1/2}$, 
${\pmb e}_{\pmb V}$ is the unit vector 
along ${\pmb V}$.

Now imagine that we add a quasiparticle of a species $i$,
of momentum ${\pmb p}$ and energy $\varepsilon_{i}({\pmb p})$,
to the system.
Then the total momentum and energy in the frame $K$ become equal to
${\pmb P}+{\pmb p}$ and $E+\varepsilon_{i}({\pmb p})$, respectively.
On the other hand, 
the momentum and energy in the frame $\overline{K}$
will be $\overline{{\pmb P}}+\overline{{\pmb p}}$
and $\overline{E}+\overline{\varepsilon}_{i}(\overline{{\pmb p}})$.
Consequently, using Eqs.\ (\ref{EEE}) and (\ref{PPP}) 
one obtains the transformation rules 
for the quasiparticle momentum and energy
\begin{eqnarray}
\varepsilon_{i}({\pmb p}) &=& 
[\overline{\varepsilon}_{i}(\overline{{\pmb p}})
+\overline{{\pmb p}}{\pmb V}] \, \gamma,
\label{quasi_energy1} \\
{\pmb p} &=& \overline{{\pmb p}} - {\pmb e_{\pmb V}} 
\left({\pmb e_{\pmb V}} \overline{{\pmb p}} 
\right)\, (1- \gamma) 
+ \overline{\varepsilon}_{i}(\overline{{\pmb p}}) 
{\pmb V}\gamma.
\label{quasi_momentum1}
\end{eqnarray}
We need to know also how the distribution function 
of quasiparticle species $i$ transforms from one frame to another.
The answer is given by the standard formula
\begin{equation}
n_i({\pmb p})=\overline{n}_i(\overline{{\pmb p}}).
\label{distrib}
\end{equation}
Assume now, that ${\pmb V}$ satisfies the inequality, $V \ll \vFa$.
Then we have also $V \ll 1$ and, 
as follows from Eqs.\ (\ref{quasi_energy1})--(\ref{distrib}),
%with the accuracy to 
keeping
linear terms in ${\pmb V}$, one gets
\begin{eqnarray}
\overline{\varepsilon}_i({\pmb p}) &=& \varepsilon_i({\pmb p}) 
+ \frac{\partial \varepsilon_i({\pmb p})}{\partial{\pmb p}} \, 
\varepsilon_i({\pmb p}) \, {\pmb V} - {\pmb p} {\pmb V},
\label{eq1}\\
\overline{n}_i({\pmb p}) &=& n_i({\pmb p}) 
+ \frac{\partial n_i({\pmb p})}{\partial {\pmb p}} \, 
\varepsilon_i({\pmb p}) \, {\pmb V}.
\label{eq2}
\end{eqnarray}
In the case of non-interacting relativistic particles 
the sum of two last terms 
on the right-hand side of Eq.\ (\ref{eq1}) equals zero, 
hence $\overline{\varepsilon}_i({\pmb p}) 
= \varepsilon_i({\pmb p})$.  

In addition to Eq.\ (\ref{eq1}), 
there is one more condition
relating $\overline{\varepsilon}_i({\pmb p})$
and $\varepsilon_i({\pmb p})$. 
In fact, it follows from Eq.\ (\ref{energy_quasi})
that for any chosen momentum ${\pmb p}$ the quasiparticle energy
$\varepsilon_i({\pmb p})$ in the frame $K$ will differ from
the energy $\overline{\varepsilon}_i({\pmb p})$ 
in the frame $\overline{K}$ 
only to the extent that 
$n_i({\pmb p})$ differs from $\overline{n}_i({\pmb p})$.
In other words,
\begin{equation}
\varepsilon_i({\pmb p}) = \overline{\varepsilon}_i({\pmb p})
+ \sum_{{\pmb p}' s' k} 
f^{ik}({\pmb p}, {\pmb p}') 
\left[ n_k({\pmb p}') -\overline{n}_k({\pmb p}')\right].
\label{sviaz}
\end{equation}
Substituting into Eq.\ (\ref{sviaz}) 
$\overline{\varepsilon}_i({\pmb p})$
and $\overline{n}_i({\pmb p})$ 
from Eqs.\ (\ref{eq1}) and (\ref{eq2}), respectively, 
one obtains
\begin{equation}
\left[ \frac{\partial \varepsilon_i({\pmb p})}{\partial{\pmb p}} \, 
\varepsilon_i({\pmb p}) - {\pmb p} \right] \, {\pmb V} 
- \sum_{{\pmb p}' s' k} f^{ik}({\pmb p}, {\pmb p}') \,
\frac{\partial n_k({\pmb p'})}{\partial {\pmb p'}} \, 
\varepsilon_k({\pmb p'}) \, {\pmb V} = 0.
\label{sviaz2}
\end{equation}
For a system in its ground state one has
$n_i({\pmb p})=n_{i0}({\pmb p})$ and
$\varepsilon_i({\pmb p})=\varepsilon_{i0}({\pmb p})$
(see Eqs.\ (\ref{Fermisphere}) and (\ref{energy0})).
At $p=\pFa$ Eq.\ (\ref{sviaz2}) relates 
the effective mass $m_i^{\ast}$ and the Landau parameters $f^{ik}_1$
\begin{equation}
\frac{\mu_i}{m_i^{\ast}} = 1 - \sum_k \frac{\mu_k G_{ik}}{n_i}.
\label{effmass}
\end{equation}
Here the number density of particle species $i$ is given by
\begin{equation}
n_i=\frac{\pFa^3}{3 \pi^2},
\label{numberdensity}
\end{equation}
while the symmetric matrix $G_{ik}$ equals
\begin{equation}
G_{ik} = \frac{1}{9 \pi^4} \, \pFa^2 \pFas^2 \, f_1^{ik}.
\label{Gik}
\end{equation}
For a liquid composed of identical particles,
Eq.\ (\ref{effmass}) transforms into the equation (13) of Ref.\ \cite{bc76}.
The non-relativistic limit of Eq.\ (\ref{effmass}) 
can be obtained if one replaces $\mu_i$ by $m_i$.
Applying then this formula to a mixture of two species of baryons,
one reproduces the result of Sj$\ddot{ {\rm o}}$berg \cite{sjoberg76} 
(see also \cite{bjk96, ch06}).

%%%%%%%%%%%%%%%%%%%%%%%%%%%%%%%%%%%%%%%%%%%%%%%%%%%%%%%%%%%%%%%%%%%%%%%
\subsection{Calculation of the relativistic entrainment matrix 
%of superfluid nucleon-hyperon mixture at zero temperature
}
\label{2b}
%%%%%%%%%%%%%%%%%%%%%%%%%%%%%%%%%%%%%%%%%%%%%%%%%%%%%%%%%%%%%%%%%%%%%%%

Let us employ the theory described above to calculate the 
relativistic entrainment matrix $Y_{ik}$
at zero temperature.
At first glance, 
%the formulae of Sec.\ IIA 
this theory
seems inappropriate for 
calculation of superfluid properties of nucleon-hyperon matter
because it describes the normal Fermi fluid. 
However, as was demonstrated by Leggett \cite{leggett65,leggett75}
in the context of superfluid $^3$He,
the particle current density ${\pmb j}_i$ of particle species $i$
in superfluid nucleon-hyperon matter
is given by the same equation
as in the case of normal (non-superfluid) matter, namely
\begin{equation}
{\pmb j}_i = \sum_{{\pmb p} s} 
\frac{\partial \varepsilon_i({\pmb p})}{\partial {\pmb p}} 
\, n_i({\pmb p}).
\label{tok}
\end{equation}
All that we need to know is how the superfluid motions modify 
the distribution function $n_i({\pmb p})$ of quasiparticles.
One also has to take into account that a change of $n_i({\pmb p})$
results in a change of the quasiparticle energy 
$\varepsilon_i({\pmb p})$. 
Leggett \cite{leggett65}
showed that this energy can be calculated 
from the same formula (\ref{energy_quasi}) as for normal matter
(see also \cite{bjk96,cj03,gh05,ch06}).

As was already mentioned in Sec.\ I,
the superfluid current 
is generated
%of particle species $i$ appear 
in the system 
when the Cooper pairs acquire a non-zero momentum $2 {\pmb Q}_i$.
In this case they are formed by pairing of quasiparticles 
with momenta $(-{\pmb p}+{\pmb Q}_i)$ and $({\pmb p}+{\pmb Q}_i)$ 
(rather than with strictly opposite momenta $-{\pmb p}$ and ${\pmb p}$, 
as it would be in the system without currents).
The distribution function $n_i({\pmb p})$ for a system with currents
can be especially easily found at zero temperature.
In this case all quasiparticles are paired and, 
up to small terms of the order of $O[(\Delta/\mu)^2+({\pmb Q}_i /\mu)^2]$
(where $\Delta$ is some characteristic value of an energy gap in 
the dispersion relation for baryons;
$\mu$ is the characteristic chemical potential of baryons),
$n_i({\pmb p})$ is a Fermi-sphere, 
shifted by the vector ${\pmb Q}_i$ in momentum space 
(see, e.g., \cite{bjk96,cj03,ch06}),
\begin{equation}
n_i({\pmb p})=\theta( \pFa- |{\pmb p}-{\pmb Q}_i|).
\label{np_shift}
\end{equation}
Here and below we assume that $Q_k \ll \pFa$.
In this case we may restrict ourselves to a linear in ${\pmb Q}_k$ terms
when calculating ${\pmb j}_i$.
Using the distribution function (\ref{np_shift}) as well as 
Eq.\ (\ref{energy_quasi}) for the energy of quasiparticle species $i$
in which $\delta n_i({\pmb p})
\equiv \theta( \pFa-|{\pmb p}-{\pmb Q}_i|)-n_{i0}({\pmb p})
\approx - \left[ \partial n_{i0}({\pmb p})/ \partial {\pmb p} \right] 
\,\,{\pmb Q}_i$, one gets from Eq.\ (\ref{tok})
\begin{equation}
{\pmb j}_i = -\sum_{{\pmb p} s} 
\frac{\partial \varepsilon_{i0}({\pmb p})}{\partial {\pmb p}} 
\, \left[ 
\frac{\partial n_{i0}({\pmb p}) 
%\theta(\pFa-|{\pmb p}|)
}{\partial {\pmb p}} 
\,{\pmb Q}_i \right]
-\sum_{{\pmb p} s} \frac{\partial}{\partial {\pmb p}} 
\left[ \sum_{{\pmb p}' s' k} f^{ik}({\pmb p}, {\pmb p}') \, 
\frac{\partial  n_{k0}({\pmb p}') }
%\theta(\pFas-|{\pmb p}'|)}
{\partial {\pmb p}'} 
\,{\pmb Q}_k \right] 
%\theta(\pFa-|{\pmb p}|) 
n_{i0}({\pmb p}).
\label{tok2}
\end{equation}
The first term in the right-hand side of Eq.\ (\ref{tok2}) equals
${\it I}= n_i/(m_i^{\ast}) \, {\pmb Q}_i$.
Integrating by parts the second term, one has
${\it II}=\sum_k G_{ik} \, {\pmb Q}_k$, 
where the matrix $G_{ik}$ is defined by Eq.\ (\ref{Gik}).
Thus, one finds for the particle current density ${\pmb j}_i$
\begin{equation}
{\pmb j}_i = \frac{n_i}{m_i^{\ast}} \, {\pmb Q}_i
+\sum_k G_{ik} \, {\pmb Q}_k. 
%+ \frac{n_i}{m_i^{\ast}} \, {\pmb Q}_i.
\label{tok3}
\end{equation}
Comparison of this result with Eq.\ (\ref{entrmatr})
allows one to determine the expression for relativistic 
entrainment matrix $Y_{ik}$ 
($\delta_{ik}$ is the Kronecker symbol)
\begin{equation}
Y_{ik} = \frac{n_i}{m_i^{\ast}} \, \delta_{ik} + G_{ik}.
\label{Yik2}
\end{equation}
%
%This matrix satisfies the condition (\ref{sumrule}) as can be checked
%with the help of the expression (\ref{effmass}) for the effective mass.
Using Eq.\ (\ref{effmass}) we verified 
that the matrix $Y_{ik}$ satisfies the condition (\ref{sumrule}).

%Calculating from Eq.\ (\ref{energy2}) 
The energy of nucleon-hyperon matter with superfluid currents
can be also expressed through the matrix $Y_{ik}$.
From Eq.\ (\ref{energy2}) it follows that
\begin{equation}
E-E_0 = \frac{1}{2} \, \sum_{i k} Y_{ik} \, {\pmb Q}_i {\pmb Q}_k.
\label{energy3} 
\end{equation}
An analogous formula, valid for an arbitrary temperature
(not only for $T=0$) was obtained for a mixture 
of two non-relativistic superfluids
by Andreev and Bashkin \cite{ab75}.
Notice that, the difference $(E-E_0)$ can be interpreted as the
energy of superfluid motion.
%This energy cannot be negative, 
For a stable superfluid ground state, $E-E_0>0$, 
and hence the quadratic form in the right-hand side 
of Eq.\ (\ref{energy3}) should be positively defined.
This leads to a set of conditions 
on the matrix $Y_{ik}$ or, equivalently,
on the Landau parameters $f_1^{ik}$. 
Here we will write only the simplest two of them
(see also \cite{ab75, ch06})
\begin{equation}
Y_{ii} \geq 0, \quad \quad Y_{ii} Y_{kk} -Y_{ik}^2 \geq 0  
\qquad (i \neq k). 
\label{Yikcond}
\end{equation}
%

%%%%%%%%%%%%%%%%%%%%%%%%%%%%%%%%%%%%%%%%%%%%%%%%%%%%%%%%%%%%%%%%%%%%%%%
\section{The relativistic entrainment matrix at zero temperature from 
the $\sigma$-$\omega$-$\rho$ model with scalar self-interactions}
\label{3}
%%%%%%%%%%%%%%%%%%%%%%%%%%%%%%%%%%%%%%%%%%%%%%%%%%%%%%%%%%%%%%%%%%%%%%%

Let us apply the general results obtained in Sec.\ II to a specific
model describing the interaction of baryons, 
the $\sigma$-$\omega$-$\rho$
mean-field model with scalar self-interactions.
Our aim will be to calculate in the frame of this model the 
relativistic entrainment matrix $Y_{ik}$ as well as 
the Landau parameters $f_l^{ik}$ of nucleon-hyperon mixture.

We choose the $\sigma$-$\omega$-$\rho$ model 
rather than, for example, more elaborated mean-field model including 
hidden strangeness $\sigma^{\ast}$ and $\phi$ mesons 
(which mediate interaction between hyperons), because of two reasons.
First of all, it is relatively simple and still realistic model to start with.
Second, the hidden strangeness mesons were originally proposed 
to simulate strong hyperon-hyperon interaction. 
However, Nagara event \cite{takahashi01} 
suggests that $\Lambda\Lambda$ interaction 
in $^6_{\Lambda \Lambda}$He can be weak.
To explain such a weak interaction, 
$\sigma$-$\omega$-$\rho$ model 
is sufficient (see, e.g., \cite{syt06}). 
Bearing this in mind and taking into account that 
the hyperon-meson coupling constants 
are known with large uncertainty, 
our choice of the model seems justifiable.

To present a quantitative example supporting 
our simple model, let us refer to a specific 
model of neutron star cores \cite{bg04}. 
This model assumes  a weak $\Lambda\Lambda$ 
interaction and includes  
the $\sigma^{\ast}$ and $\phi$ mesons. 
As seen in Fig. 6 of 
Ref.\ \cite{bg04}, the contribution of the 
$\sigma^{\ast}$ meson to the hyperon effective masses
is at most a few percent of that resulting from 
the $\sigma$ meson. Similarly, the $\phi$ meson 
potential in which hyperons move is at most 
a few percent of the contribution resulting 
from the $\omega$ meson. This example 
suggests that the contribution of the 
$\sigma^{\ast}$ meson to the hyperon entrainment
matrix is small, while that of the 
$\phi$ meson may be expected to be small.

%%%%%%%%%%%%%%%%%%%%%%%%%%%%%%%%%%%%%%%%%%%%%%%%%%%%%%%%%%%%%%%%%%%%%%%
\subsection{$\sigma$-$\omega$-$\rho$ mean-field model
with scalar self-interactions: 
general equations}
\label{3a}
%%%%%%%%%%%%%%%%%%%%%%%%%%%%%%%%%%%%%%%%%%%%%%%%%%%%%%%%%%%%%%%%%%%%%%%
The $\sigma$-$\omega$-$\rho$ model with scalar self-interactions 
is described in detail in the monograph 
by Glendenning \cite{glendenning00} (see also \cite{glendenning85}).
Here we briefly discuss its main equations 
which will be used below to calculate
the relativistic entrainment matrix $Y_{ik}$.
%Unless otherwise is stated, throughout the paper we use the system of units
%in which the Planck constant $\hbar$, the Boltzmann constant $k_{\rm B}$, 
%and the speed of light $c$ equal unity, 
%$\hbar=k_{\rm B}=c=1$.
%We also imply that the subscripts $i$ and $k$ refer to baryons.
%$i$, $k= n, p, \Sigma,$ and $\Lambda$. 
Let us consider a system of baryons 
$n$, $p$, $\Lambda$, and $\Sigma$
in some uniform state.
%We are mainly interested in the baryons of four species: 
%neutrons ($n$), protons ($p$), $\Sigma^{-}$, 
%and $\Lambda$-hyperons ($\Sigma$ and $\Lambda$, respectively).
Interactions among those baryons 
are mediated by three different kinds of meson fields:
scalar $\sigma$-field, vector $\omega$-field 
and an isospin triplet of charged vector ${\vec \rho}$-fields.
The mean-field approximation assumes that the 
$\sigma$-, $\omega$-, and ${\vec \rho}$-fields are replaced
by their mean expectation values in the chosen state.
We denote these values by $\sigma$, $\omega^{\mu}$, and 
${\vec \rho}^{\, \mu}=(\rho_1^{\mu},\rho_2^{\mu},\rho_3^{\mu})$, 
respectively ($\mu$ is the space-time index).
These mean values are to be calculated from the following 
(averaged) Euler-Lagrange equations \cite{glendenning00}
\begin{eqnarray}
m_{\sigma}^2 \sigma &=& -b m_n \, g_{\sigma n} \, (g_{\sigma n} \sigma)^2
-c g_{\sigma n} \, (g_{\sigma n} \sigma)^3
\nonumber \\
%&+& \sum_i \frac{2 s_i+1}{(2 \pi)^3} \, g_{\sigma i} \, 
%\int
%\frac{m_i-g_{\sigma i} \sigma}{\sqrt{p^2+(m_i-g_{\sigma i} \sigma)^2} } 
%\, n_i({\pmb p}) \, \dd {\pmb p},
%
&+& \sum_{{\pmb p} s i} \, g_{\sigma i} \, 
\frac{m_i-g_{\sigma i} \sigma}
{\sqrt{({\pmb p}-g_{\omega i} \, {\pmb \omega}
-g_{\rho i} \, I_{3i} \, {\pmb \rho}_3)^2
+(m_i-g_{\sigma i} \sigma)^2} } 
\, n_i({\pmb p}),
\label{sigma} \\
\omega^{\mu}&=& \sum_i \frac{g_{\omega i}}{m^2_{\omega}} \,\, j^{\mu}_i,
\label{omega}\\
\rho_{1}^{\mu}&=&\rho_{2}^{\mu}=0,
\label{rho12}\\
\rho^{\mu}_{3}&=& \sum_i \frac{g_{\rho i}}{m^2_{\rho}} \,\, I_{3i} \, j^{\mu}_i.
\label{rho3}
\end{eqnarray}
One sees that only the third isospin component $\rho_3^{\mu}$ 
of the ${\vec \rho}$-field, which corresponds to the neutral rho meson,
has non-zero mean value. 
In Eqs.\ (\ref{sigma})--(\ref{rho3}) 
the summation is performed over the baryon species $i=n$, $p$, $\Lambda$, 
and $\Sigma$; 
$m_{l}$ is the mass of meson species $l=\sigma$, $\omega$, or $\rho_{1,2,3}$;
$g_{li}$ is the coupling constant of meson $l$ and baryon $i$; 
%$m_i$, $s_i$, and 
$I_{3i}$ is the isospin projection for baryon species $i$.
Furthermore, $n_i({\pmb p})$ is (as in Sect.\ IIA) 
a distribution function of 
particle species $i$; $b$ and $c$ are some dimensionless constants 
describing self-interaction of 
the scalar $\sigma$-field;
${\pmb \omega}$ and ${\pmb \rho}_3$ are the spatial components of 
four-vectors $\omega^{\mu}$ and $\rho_3^{\mu}$, respectively.
The $\omega$- and $\rho_3$-fields are 
generated by the baryon four-currents $j^{\mu}_i$ 
on the right-hand side of Eqs.\ (\ref{omega}) and (\ref{rho3}).
They are given by
\begin{eqnarray}
j^0_i &=& n_i=
%\frac{2 s_i+1}{(2 \pi)^3} \int 
\sum_{{\pmb p} s} n_i({\pmb p}), 
% \, \dd {\pmb p},
\label{j0}\\
{\pmb j}_i&=& 
%\frac{2 s_i+1}{(2 \pi)^3} \int 
\sum_{{\pmb p} s} \frac{\partial E_{i}({\pmb p})}{\partial {\pmb p}} \,
n_i({\pmb p}),
%\, \dd {\pmb p},
\label{j}
\end{eqnarray}
where the number density $n_i$ and the particle current density ${\pmb j}_i$ 
are measured in the laboratory frame;
$E_i({\pmb p})$ is the energy of a baryon species $i$
\begin{equation}
E_i({\pmb p})=g_{\omega i} \, \omega^0 + g_{\rho i} \, I_{3i} \, \rho^0_3 +
\sqrt{({\pmb p}-g_{\omega i} \, {\pmb \omega}- g_{\rho i} \, I_{3i} \, {\pmb 
\rho}_{3})^2 
+ (m_i-g_{\sigma i} \sigma)^2}.
\label{energy}
\end{equation}
In Eqs.\ (\ref{sigma}), (\ref{j0}), and (\ref{j}) 
the summation is performed over the momentum states occupied by the particles.
If our system is not only uniform but also isotropic 
%(and temperature equals zero, $T=0$)
then (at zero temperature) 
the distribution function $n_i({\pmb p})$ is a Fermi sphere centered 
at ${\pmb p}=0$ in the momentum space, so that we have 
(see Eq.\ (\ref{Fermisphere}))
\begin{equation}
n_i({\pmb p})=n_{i0}({\pmb p}). 
%= \theta(\pFa-p). 
\label{np}
\end{equation}
%
%Here $\theta(x)$ is the step function, $\theta(x)=1$ if $x \leq 0$ and 0 
%otherwise;
%$p_{{\rm F}i}$ is the Fermi momentum of particle species $i$. 
%In view of Eqs.\ (\ref{j0}) and (\ref{np}) 
%it is related to the number density $n_i$ by the following equation
Substituting the distribution function (\ref{np}) into 
Eq.\ (\ref{j0}) one obtains that the time component 
$j_i^0=n_i$ 
%of $j_i^{\mu}$ 
is given by Eq.\ (\ref{numberdensity}).
%
%\begin{equation}
%j^0_i = n_i= \frac{\pFa^3}{3 \pi^2}.
%\label{j0i}
%\end{equation}
%
Moreover, in this special case the spatial components 
of four-vectors $\omega^{\mu}$, 
$\rho^{\mu}_3$, and $j^{\mu}_i$ vanish, 
${\pmb \omega}={\pmb \rho}_3={\pmb j}_i=0$ 
(there is no preferred direction!), 
while $\sigma$-field and the time components are still given by
Eqs.\ (\ref{sigma}), (\ref{omega}), and (\ref{rho3}) with 
$n_i(\pmb p)$ and $j^0_i$ taken from Eqs.\ (\ref{np}) and (\ref{numberdensity}), 
respectively.
The chemical potential $\mu_i$ of baryon species $i$ is presented in the form
\begin{equation}
\mu_i=g_{\omega i} \, \omega^0 + g_{\rho i} \, I_{3i} \, \rho^0_3 +
\sqrt{p_{{\rm F}i}^2 + (m_i-g_{\sigma i} \sigma)^2}.
\label{mu}
\end{equation}
It is the energy of a particle on the Fermi surface.

%%%%%%%%%%%%%%%%%%%%%%%%%%%%%%%%%%%%%%%%%%%%%%%%%%%%%%%%%%%%%%%%%%%%%%%
\subsection{The relativistic entrainment matrix from the 
$\sigma$-$\omega$-$\rho$ mean-field model}
\label{3b}
%%%%%%%%%%%%%%%%%%%%%%%%%%%%%%%%%%%%%%%%%%%%%%%%%%%%%%%%%%%%%%%%%%%%%%%

A derivation of the matrix $Y_{ik}$ in the frame of the $\sigma$-$\omega$-$\rho$
mean-field model with scalar self-interactions is completely analogous 
to the derivation presented in Sec.\ IIB 
for the case of relativistic Landau Fermi-liquid theory.
In nucleon-hyperon matter in which the superfluid currents are generated,
the distribution function for baryon species $i$ 
is approximately described by Eq.\ (\ref{np_shift}).

The superfluid current density ${\pmb j}_i$ is given by Eq.\ (\ref{j})
with the energy $E_i({\pmb p})$ calculated from Eq.\ (\ref{energy})
with the help of Eqs.\ (\ref{sigma}), (\ref{omega}), and (\ref{rho3}).

As it was mentioned in Sec.\ IIB,
we restrict ourselves to a linear approximation when
calculating ${\pmb j}_i$ as a function of ${\pmb Q}_k$.
In this approximation the scalar $\sigma$-field as well as 
the time components $\omega^0$ and $\rho_3^0$ remain the same
(their variation $\sim {\pmb Q}_i {\pmb Q}_k$), 
whereas the spatial components ${\pmb \omega}$ and ${\pmb \rho}_3$ 
depend on some linear combinations of the vectors ${\pmb Q}_k$.
It follows from Eqs.\ (\ref{j}) that
\begin{eqnarray}
{\pmb j}_i &=&
%\frac{2 s_i+1}{(2 \pi)^3} \int
\sum_{{\pmb p} s } \frac{\partial E_{i}({\pmb p})}{\partial {\pmb p}} \,
\, \theta(\pFa - |{\pmb p}-{\pmb Q}_i|) 
%\,  \dd {\pmb p}
\nonumber \\
&=& \sum_{{\pmb p} s } 
\frac{\partial E_{i}({\pmb p}+{\pmb Q}_i)}{\partial {\pmb p}} \,
\, n_{i0}({\pmb p})
\nonumber\\ 
&=& \sum_{{\pmb p} s } 
\frac{\partial}{\partial {\pmb p}} \left[
\sqrt{p^2+(m_i-g_{\sigma i} \sigma)^2} 
+ \frac{ {\pmb p} \left( {\pmb Q}_i -g_{\omega i} {\pmb \omega} 
- g_{\rho i} I_{3i} {\pmb \rho}_3  \right) }
{ \sqrt{p^2+(m_i-g_{\sigma i} \sigma)^2}   }
\right] n_{i0}({\pmb p})
\nonumber\\
&=& \frac{n_i}{ \sqrt{p^2_{{\rm F}i}+(m_i-g_{\sigma i} \sigma)^2} } \, 
\left( {\pmb Q}_i -g_{\omega i} {\pmb \omega} 
- g_{\rho i} I_{3i} {\pmb \rho}_3  \right).
\label{jj}
\end{eqnarray}
This equation should be supplemented by the expressions 
(\ref{omega}) and (\ref{rho3}) for ${\pmb \omega}$ and ${\pmb \rho}_3$,
respectively
\begin{eqnarray}
{\pmb \omega}  &=& \sum_i \frac{g_{\omega i}}{m^2_{\omega}} \,\, {\pmb j}_i,
\label{omega2}\\
{\pmb \rho}_{3}&=& \sum_i \frac{g_{\rho i}}{m^2_{\rho}} \,\, I_{3i} \, 
{\pmb j}_i.
\label{rho32}
\end{eqnarray}
Solving the system of six equations (\ref{jj})--(\ref{rho32})
one can find ${\pmb j}_i$ and the vectors 
${\pmb \omega}$ and ${\pmb \rho}_3$ as functions of ${\pmb Q}_k$.
In this way the relativistic entrainment matrix $Y_{ik}$
can be determined at zero temperature.
The analytic expression for $Y_{ik}$ is given in Appendix.
It is easy to verify that the matrix $Y_{ik}$ satisfies the condition 
(\ref{sumrule}).
%
%%%%%%%%%%%%%%%%%%%%%%%%%%% FIGURE 1  %%%%%%%%%%%%%%%%%%%%%%%%%%%%%
\begin{figure}[t]
\setlength{\unitlength}{1mm}
\leavevmode
\hskip  0mm
\includegraphics[width=150mm,bb=15  300  550  550,clip]{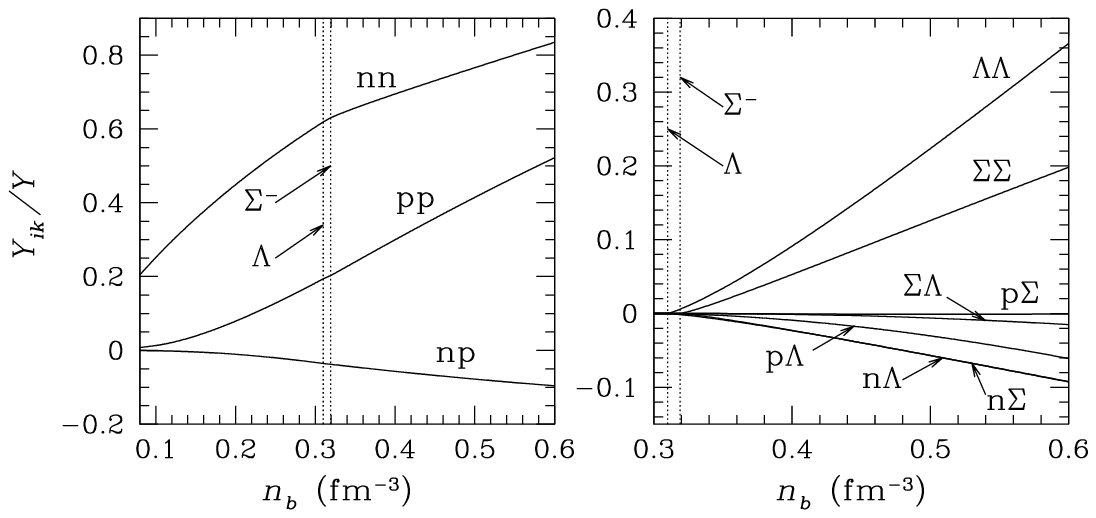}
\caption
{
Normalized symmetric matrix $Y_{ik}/Y$ 
as a function of $n_b$ for the third equation 
of state of Glendenning \cite{glendenning85}.
The normalization constant $Y=3 n_0/\mu_{n}(3 n_0) 
= 2.48 \times 10^{41}$ erg$^{-1}$ cm$^{-3}$.
Solid lines show the elements of the matrix $Y_{ik}/Y$;
each curve is marked by the corresponding symbol $ik$
($i, k = n$, $p$, $\Lambda$, $\Sigma$). 
Vertical dotted lines indicate the thresholds for the
appearance of (from left to right)  
$\Lambda$- and $\Sigma^{-}$-hyperons.
}
\label{1fig}   
\end{figure}
%%%%%%%%%%%%%%%%%%%%%%%%%%%%%%%%%%%%%%%%%%%%%%%%%%%%%%%%%%%%%%%%
%

%Note that, in the limiting case of nucleon matter 
%our results for the matrix $Y_{ik}$ do not reproduce 
%those  of Ref.\ \cite{cj03}.
%
Note that, in the limiting case 
considered by Comer and Joynt \cite{cj03}
our results for the matrix $Y_{ik}$ 
do not reproduce theirs. 
Their results do not satisfy 
the condition (\ref{sumrule}).
%
%we failed to reproduce 
%their results for $Y_{ik}$.
%
Let us remind that the authors of Ref.\ \cite{cj03}
considered {\it asymmetric} nuclear matter composed of 
neutrons, protons, and electrons. 
They assumed that nucleons interact through 
$\sigma$- and $\omega$-fields 
(the neutral $\rho_3$-field and self-interactions 
of $\sigma$-field were neglected).
The criticism of such 
assumption can be found in Ref.\ \cite{chamel08}.

Fig.\ 1 presents the normalized elements $Y_{ik}/Y$ 
of symmetric matrix $Y_{ik}$, 
calculated using Eq.\ (\ref{Yik3}),
as functions of the baryon number density
$n_{b} = n_n + n_p + n_{\Sigma} + n_{\Lambda}$ 
for the {\it third} equation of state 
of Glendenning \cite{glendenning85}.
The constant $Y$ equals
$Y=3 n_0/\mu_{n}(3 n_0) = 2.48 \times 10^{41}$ 
erg$^{-1}$ cm$^{-3}$, 
where $n_0=0.16$ fm$^{-3}$ is the 
%number density 
%of nucleons in atomic nuclei;
normal nuclear density;
$\mu_{n}(3 n_0) = 1.94 \times 10^{-3}$ erg 
is the neutron chemical potential at $n_b =3 n_0$.
Each curve on the figure is plotted 
for some normalized element of the matrix $Y_{ik}$
and marked with two particle species indices $ik$.
For instance, symbols $n \Lambda$ on the figure (right panel)
mark a curve plotted for the element $Y_{n \Lambda}/Y$ 
($=Y_{\Lambda n}/Y$).
The chosen equation of state predicts first the appearance of $\Lambda$-hyperons
at $n_b = n_{b \Lambda}=0.310$ fm$^{-3}$ and then $\Sigma^-$-hyperons
at $n_b=n_{b \Sigma}=0.319$ fm$^{-3}$. 
One sees that at $n_b<n_{b \Lambda}$ (no hyperons)
all the components of $Y_{ik}$, related to hyperons, become zero.

%%%%%%%%%%%%%%%%%%%%%%%%%%%%%%%%%%%%%%%%%%%%%%%%%%%%%%%%%%%%%%%%%%%%%%%
\subsection{Calculation of Landau parameters}
\label{3c}
%%%%%%%%%%%%%%%%%%%%%%%%%%%%%%%%%%%%%%%%%%%%%%%%%%%%%%%%%%%%%%%%%%%%%%%
The $\sigma$-$\omega$-$\rho$ 
model described above can be 
reformulated in terms of the relativistic 
Landau theory of Fermi liquids (see Sec.\ II).
For that, it is necessary to calculate the Landau parameters 
of nucleon-hyperon matter.
In case of nucleon matter the Landau parameters
were calculated for various
%modifications of the 
relativistic mean-field models in a series of papers 
(see, e.g., \cite{matsui81,hm87,cgl01,cgl02,cgl03,pbd08})
The derivation of these parameters for nucleon-hyperon matter
is quite similar.
The main idea of the derivation is 
to consider a small deviation of 
the distribution function of baryon species $i$ 
from $n_{i0}({\pmb p})$ (see Eq.\ \ref{Fermisphere})
and to analyze how it modifies the energy of baryon species $k$.
Then the result should be compared with the corresponding 
Eq.\ (\ref{energy_quasi}) for the energy variation
in the frame of the Landau theory.
In this way one obtains the function $f^{ik}({\pmb p},{\pmb p}')$
or, equivalently, the parameters $f^{ik}_l$.
In Refs.\ \cite{matsui81,hm87,cgl01,cgl02,cgl03,pbd08},  
dealing with the case of nucleon matter, it is shown that
%(regardless of a mean-field model employed) 
only first two Landau parameters 
are non-zero: $f^{ik}_0$ and $f^{ik}_1$. 
We checked that the same is true for nucleon-hyperon matter,
$f^{ik}_l=0$ at $l \geq 2$.
In view of this observation it is enough to find 
only the parameters $f^{ik}_0$ and $f^{ik}_1$.

Strictly speaking, the parameters $f_1^{ik}$ have already 
been calculated in the previous section.
Indeed, it follows from Eq.\ (\ref{Yik2}) that
%using Eq.\ (\ref{Yik3}) for the matrix $Y_{ik}$
%together with Eq.\ (\ref{Yik2}), one finds
%
\begin{equation}
f_1^{ik} = \frac{9 \pi^4}{\pFa^2 \pFas^2} \, 
\left( Y_{ik} - \frac{n_i}{m_i^{\ast}} \, \delta_{ik}
\right),
\label{f1ik}
\end{equation}
where $Y_{ik}$ is given by Eq.\ (\ref{Yik3}) and
the Landau effective masses $m_i^{\ast}$ 
(not to be confused with the Dirac effective mass!)
equal
\begin{equation}
m_i^{\ast}=\frac{\pFa}{\left| \partial E_i({\pmb p})/\partial {\pmb p} 
\right|}_{p=\pFa}
=\sqrt{\pFa^2+(m_i-g_{\sigma i} \sigma)^2}.
\label{effmassLandau}
\end{equation}
Fig.\ 2 illustrates the dependence of normalized 
Landau effective mass $m_i^{\ast}/m_i$ ($i=n$, $p$, $\Lambda$, $\Sigma$)
on $n_b$ for the third equation of state of Glendenning \cite{glendenning85}.
%
%%%%%%%%%%%%%%%%%%%%%%%%%%% FIGURE 2  %%%%%%%%%%%%%%%%%%%%%%%%%%%%%
\begin{figure}[t]
\setlength{\unitlength}{1mm}
\leavevmode
\hskip  0mm
\includegraphics[width=85mm,bb=18  145  580  690,clip]{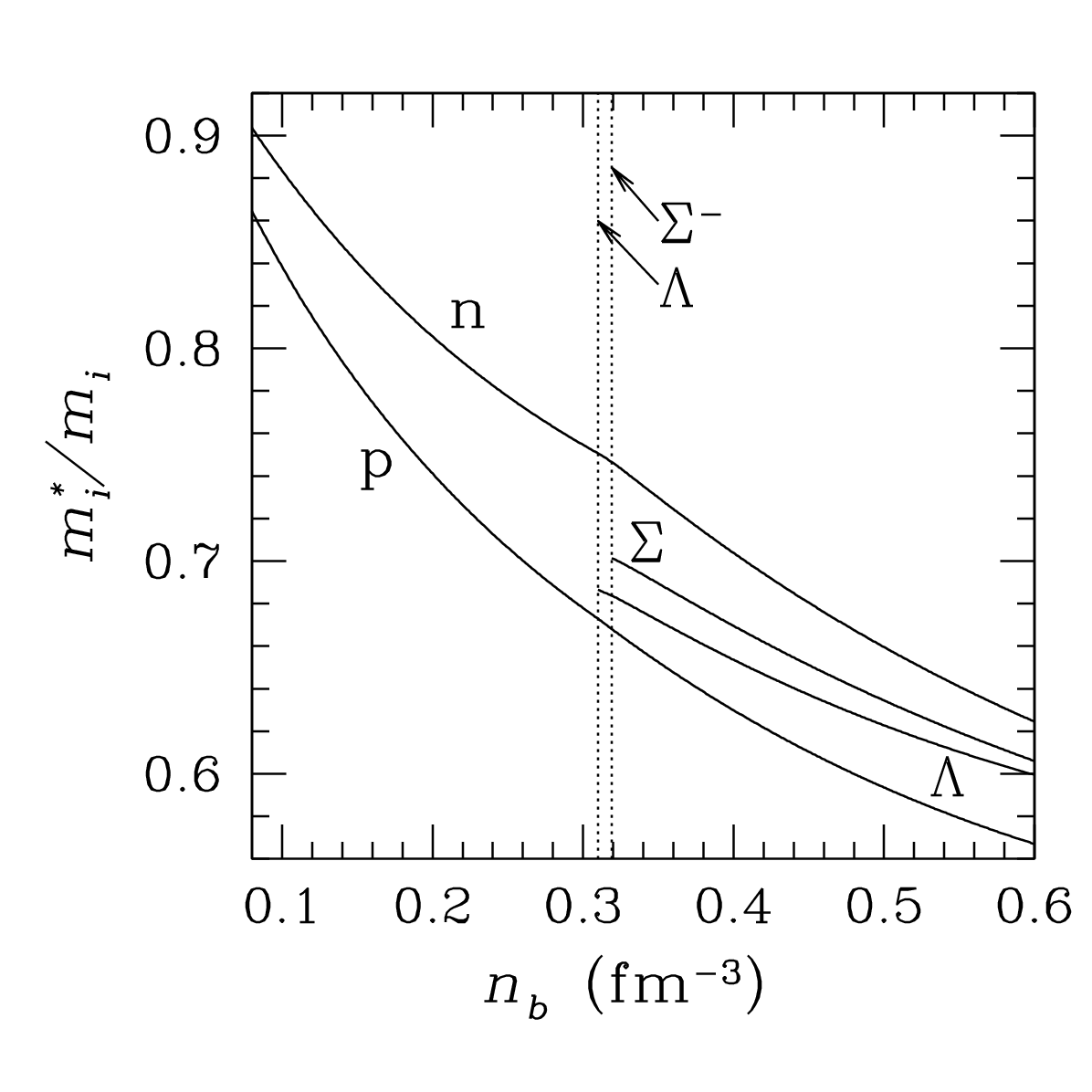}
\caption
{
The normalized Landau effective masses $m^{\ast}_i/m_i$ 
($i=n$, $p$, $\Lambda$, $\Sigma$) versus $n_b$
for the third equation of state of Ref.\ \cite{glendenning85}.
Vertical dotted lines indicate thresholds for the appearance of 
(from left to right) $\Lambda$- and $\Sigma^{-}$-hyperons.
}
\label{2fig}   
\end{figure}
%%%%%%%%%%%%%%%%%%%%%%%%%%%%%%%%%%%%%%%%%%%%%%%%%%%%%%%%%%%%%%%%
%

Now let us calculate the parameters $f^{ik}_0$. 
For that we slightly vary the Fermi momentum $\pFa$
by a small quantity $\Delta \pFa$. 
This will alter $n_{i0}({\pmb p})$ by
\begin{equation} 
\delta n_i({\pmb p})
=\theta(\pFa + \Delta \pFa - p) - n_{i0}({\pmb p}), 
%\approx - \frac{\partial n_{i0}({\pmb p})}{\partial p} \, \Delta \pFa,
\label{delta_ni}
\end{equation}
while the variation of the energy of baryon species $i$ 
(on the Fermi surface)
will be (see Eq.\ (\ref{energy_quasi}))
\begin{equation}
\delta \varepsilon_i(\pFa)= \sum_k f_0^{ik} \, \delta n_k.
\label{delta_e}
\end{equation}
Here $\delta n_k=\pFas^2 \, \Delta \pFas/\pi^2$ 
is the variation of the number density of particle species $i$. 
On the other hand, if we consider 
the $\sigma$-$\omega$-$\rho$ model,
the variation of the baryon energy  
on the Fermi surface 
will be (in the first approximation, 
see Eq.\ (\ref{energy})) 
%or (\ref{mu}))
%
\begin{equation}
\delta E_i(\pFa)= g_{\omega i} \, \delta \omega^0 
+ g_{\rho i} I_{3i} \, \delta \rho_3^0 
%+ \left. \frac{\partial E_i({\pmb p})}{\partial \sigma} 
%\right|_{p=\pFa} 
- \frac{g_{\sigma i} (m_i-g_{\sigma i} \sigma)} {m^{\ast}_i}
\delta \sigma,
\label{energy_change}
\end{equation}
The small terms $\delta \sigma$, 
$\delta \omega^0$, and $\delta \rho_3^0$
can be expressed through $\delta n_k$ from Eqs.\ (\ref{sigma}), 
(\ref{omega}), and (\ref{rho3}), respectively
\begin{eqnarray}
\delta \sigma &=& \frac{1}{L(\sigma)} \, \sum_k
\frac{g_{\sigma k} \, (m_k - g_{\sigma k} \sigma)}{m_k^{\ast}} 
\, \delta n_k,
\label{ds} \\
\delta \omega^0 &=& \sum_k \frac{g_{\omega k}}{m_{\omega}^2} \, \delta n_k,
\label{do}\\
\delta \rho_3^0 &=& \sum_k \frac{g_{\rho k}}{m_{\rho}^2} 
\, I_{3k} \, \delta n_k.
\label{dr}
\end{eqnarray}
The function $L(\sigma)$ in Eq.\ (\ref{ds}) is given by
\begin{eqnarray}
L(\sigma) &=& 
\frac{\partial}{\partial \sigma} \left[  
m_{\sigma}^2 \sigma + b m_n \, g_{\sigma n} \, (g_{\sigma n} \sigma)^2
+c g_{\sigma n} \, (g_{\sigma n} \sigma)^3 \right.
\nonumber \\
&-& \left. \sum_{{\pmb p} s i} \, 
\frac{g_{\sigma i} (m_i-g_{\sigma i} \sigma)}{\sqrt{p^2+(m_i-g_{\sigma i} 
\sigma)^2} } 
\, n_{i0}({\pmb p}) \right].
\label{Lsigma}
\end{eqnarray}
Substituting now Eqs.\ (\ref{ds})--(\ref{dr}) into 
Eq.\ (\ref{energy_change}) and comparing 
the resulting expression with Eq.\ (\ref{delta_e}), 
one finds the Landau parameters $f_0^{ik}$
\begin{equation}
f_0^{ik} = \frac{g_{\omega i} g_{\omega k}}{m_{\omega}^2} 
+ \frac{g_{\rho i} I_{3i} \, g_{\rho k} I_{3k}}{m_{\rho}^2}
- \frac{1}{L(\sigma)} 
\frac{g_{\sigma i} (m_i-g_{\sigma i} \sigma)}{m_i^{\ast}} \,
\frac{g_{\sigma k} (m_k-g_{\sigma k} \sigma)}{m_k^{\ast}}.
\label{f0ik}
\end{equation}
%
%%%%%%%%%%%%%%%%%%%%%%%%%%% FIGURE 3  %%%%%%%%%%%%%%%%%%%%%%%%%%%%%
\begin{figure}[t]
\setlength{\unitlength}{1mm}
\leavevmode
\hskip  0mm
\includegraphics[width=150mm,bb=30  300  550  550,clip]{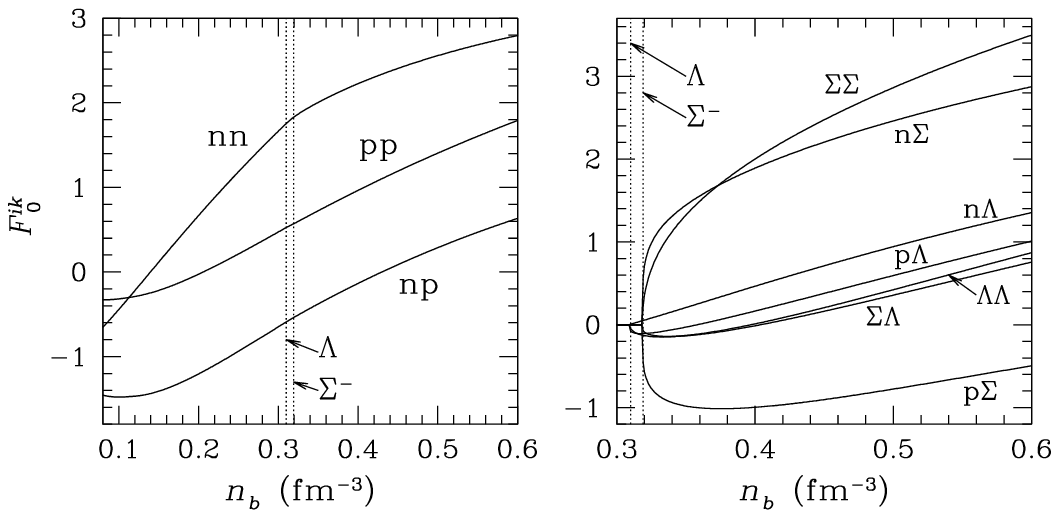}
\caption
{
Dimensionless Landau parameters $F_0^{ik}$ versus $n_b$
for the third equation of state of Ref.\ \cite{glendenning85}.
%Nucleon matter is unstable at 
%$n_b \leq 0.34 n_0 \approx 0.055$ fm$^{-3}$ (shaded region).
Other notations are the same as in Figs.\ 1 and 2.
}
\label{3fig}   
\end{figure}
%%%%%%%%%%%%%%%%%%%%%%%%%%%%%%%%%%%%%%%%%%%%%%%%%%%%%%%%%%%%%%%%
%
%Notice that, 
%as it should be, 
It follows from Eqs.\ (\ref{f1ik}) and (\ref{f0ik})
that the parameters $f_0^{ik}$ and $f_1^{ik}$ are indeed symmetric 
in the indices $i$ and $k$.

Just as the parameters $f_1^{ik}$ must guarantee the positive 
definiteness of the quadratic form (\ref{energy3}),
the parameters $f_0^{ik}$ must satisfy a number of conditions.
These conditions 
are related to stability of
charged multi-component mixture with 
respect to density fluctuations
and were carefully analysed 
for nucleon matter 
(see, e.g., \cite{bbp71, prl95, bcdl98, dmc08, dpsbc08}).
They depend essentially on the matter composition
and on the applied perturbation.
Here we consider an equilibrated matter 
of massive neutron stars 
composed not only of nucleons ($n$ and $p$) 
and hyperons ($\Lambda$ and $\Sigma$) 
but also of electrons ($e$) and muons ($\mu$).
As an example, we analyse the stability of such matter
with respect to long-wavelength density fluctuations.

The stability conditions follow from the requirement 
of minimum of the free energy
$F \equiv E-\sum_j \mu_j n_j$ (at fixed $\mu_j$;
$j=n$, $p$, $\Lambda$, $\Sigma$, $e$, $\mu$) 
for the system in thermodynamic equilibrium, at $T=0$.
Using Eq.\ (\ref{energy2}) for the variation 
of energy of baryons, it is easy
to find a variation 
$\delta F = \delta E - \sum_j \mu_j \, \delta n_j$
caused by a small change of $\delta n_j({\pmb p})$ 
%(see Eq.\ (\ref{delta_ni})),
(see Eq.\ (\ref{delta_ni})
with $j$ instead of $i$):
\begin{equation}
\delta F = \frac{1}{2} \sum_{ik} 
\left( 
%\frac{\pi^2}{m_i^{\ast} \pFa} 
\frac{1}{N_i}
\delta_{ik} + f_0^{ik} \right)
\delta n_i \, \delta n_k + \frac{1}{2} \, 
\frac{\partial \mu_e}{\partial n_e} \, (\delta n_e)^2
+ \frac{1}{2} \, 
\frac{\partial \mu_\mu}{\partial n_\mu} \, (\delta n_\mu)^2.
\label{dF}
\end{equation}
Here $N_i \equiv m_i^{\ast} \pFa/\pi^2$ is the 
density of states of particle species $i$ on the Fermi surface;
$\mu_l$ and $n_l$ are, respectively, the relativistic chemical potential 
and number density of electrons ($l=e$) and muons ($l=\mu$).
To derive Eq.\ (\ref{dF}) we presented the variation 
$\delta E_l$ of the energy $E_l$ of leptons, 
in the form ($l=e$, $\mu$)
\begin{equation}
\delta E_l = \frac{\partial E_l}{\partial n_l} 
\, \delta n_l + \frac{1}{2} \, 
\frac{\partial^2 E_l}{\partial n_l^2} 
\, (\delta n_l)^2 = \mu_l \, \delta n_l 
+ \frac{1}{2} \, 
\frac{\partial \mu_l}{\partial n_l} 
\, (\delta n_l)^2.
\label{dEl}
\end{equation}
As it should be, the expansion of $F$ begins with the terms
of the second order in $\delta n_j$.
The requirement of minimum of $F$ means that $\delta F \geq 0$,
that is the quadratic form in the right-hand side of 
Eq.\ (\ref{dF}) must be positively defined.

In Eq.\ (\ref{dF}) for the variation $\delta F$ 
of the free energy, we neglected a positive term 
related to the Coulomb energy of the perturbed matter.
However, it must be taken into account
if the perturbed matter acquired a non-zero charge, 
which is the case when $\delta n_p - \delta n_e 
-\delta n_\mu - \delta n_{\Sigma} \neq 0$.
The contribution of the Coulomb energy 
to $\delta F$ is then $\sim q^{-2}$ 
(see, e.g., \cite{bbp71, prl95, dmc08}),
where $q$ is the wave number of plane-wave density fluctuation.
Here we are interested only in the limit of
long wavelengths, for which $q \rightarrow 0$.
In this limit, the positive Coulomb energy 
can be arbitrarily large, 
so that the matter is stable
against the long-wavelength density perturbations
at {\it any density}.
To exclude the `stabilizing' contribution 
of the Coulomb energy
%from our consideration
we 
%have to 
consider only those variations $\delta n_j$
of the number densities 
which preserve the charge neutrality, 
\begin{equation}
\delta n_p - \delta n_e 
-\delta n_\mu - \delta n_{\Sigma} =0.
\label{quasi}
\end{equation}
Expressing $\delta n_e$ using this equation and
substituting it into Eq.\ (\ref{dF}), one finds
\begin{equation}
\delta F = \frac{1}{2} \sum_{jm} A_{jm} 
\, \delta n_j \, \delta n_m,
\label{dF1}
\end{equation}
where the indices $j$ and $m$ run over all particle species
except for electrons.
The $5 \times 5$ matrix $A_{jm}$ is given by
\begin{equation}
A_{jm} = \left( 
\frac{\delta_{jm}}{N_j} + f^{jm}_0 \right)
\delta_{jb} \delta_{mb} 
%\nonumber \\
%&&  
+\frac{\partial \mu_e}{\partial n_e}  \, q_j q_m
%+ \frac{\partial \mu_e}{\partial n_e}
%(\delta_{jp} \delta_{mp} + \delta_{j \Sigma} \delta_{m \Sigma}
%+ \delta_{j \mu} \delta_{m \mu} 
%-\delta_{j \Sigma} \delta_{m p}-\delta_{j p} \delta_{m \Sigma}
%- \delta_{j p} \delta_{m \mu} - \delta_{j \mu} \delta_{m p} 
%+ \delta_{j \Sigma} \delta_{m \mu}+ \delta_{j \mu} \delta_{m \Sigma})
%\nonumber\\
%&& + 
+\frac{\partial \mu_\mu}{\partial n_\mu} \, 
\delta_{j \mu} \delta_{m \mu}. 
%\frac{\partial \mu_\mu}{\partial n_\mu}.
\label{Ajm}
\end{equation}
Here $\delta_{j b}$ and $\delta_{m b}$ equal 1 if 
$j$ and $m=n$, $p$, $\Lambda$, $\Sigma$ and 0 otherwise;
$q_j$ and $q_m$ are, respectively, the electric charges 
of particle species $j$ and $m$ in units 
of proton charge (e.g., $q_e=-1$).

The requirement of positive definiteness 
of the quadratic form (\ref{dF1})
imposes a set of conditions 
on the matrix elements $A_{jm}$ or, equivalently, 
on the parameters $f_0^{ik}$;
we write out only the simplest two of them
\begin{eqnarray}
&&
%\frac{\pi^2}{m_i^{\ast} \pFa} 
%\frac{1}{N_i}
%+ f_0^{ii} \geq 0,
A_{jj} \geq 0,
\label{usl1}\\
%&&\left(
%%\frac{\pi^2}{m_i^{\ast} \pFa} 
%\frac{1}{N_i}
%+ f_0^{ii} \right) 
%\left(
%%\frac{\pi^2}{m_k^{\ast} \pFas} 
%\frac{1}{N_k}
%+ f_0^{kk} \right) 
%- \left( f_0^{ik} \right)^2 \geq 0  
&& A_{jj} A_{mm} - \left(A_{jm} \right)^2 \geq 0
\qquad (j \neq m).
\label{usl2}
\end{eqnarray}
These conditions are very well known in the literature
devoted to stability of nucleon matter 
(see, e.g., \cite{bbp71, prl95, dmc08}).
%The condition (\ref{usl1}) is very well known 
%in the Landau theory describing 
%liquid composed of identical particles \cite{lp80}.
For a mixture composed 
of {\it neutral} strongly interacting 
baryons they can be simplified and presented in the form 
(see, e.g., \cite{bcdl98})
\begin{eqnarray}
&&
1+ F_0^{ii} \geq 0,
\label{usl11}\\
&& \left(1+F^{ii}_0 \right) \left(1+F^{kk}_0 \right) 
- \left(F^{ik}_0 \right)^2 \geq 0 \qquad (i \neq k),
\label{usl22}
\end{eqnarray}
where the indices $i$ and $k$ refer to baryons 
and we introduced the dimensionless Landau parameters $F_l^{ik}$,
\begin{equation}
F_{l}^{ik} \equiv 
%\sqrt{  \frac{m_i^{\ast} \pFa}{\pi^2} }\, 
%\sqrt{\frac{m_k^{\ast} \pFas}{\pi^2} } \, 
\sqrt{N_i N_k} \,
f_l^{ik}.
\label{Fl_ik}
\end{equation}
%
%it can be rewritten in the 
%familiar form, $F_0^{ii} \geq -1$.

Our results are illustrated in Figs.\ 3 and 4,
where the parameters $F_0^{ik}$ 
and $F_1^{ik}$ are presented for the third equation of state 
of Glendenning \cite{glendenning85} as functions of $n_b$.
The Landau parameters for neutrons and protons are plotted
on the left panel in Figs.\ 3 and 4 ($i$, $k$= $n$, $p$).
The right panel demonstrates the Landau parameters related to hyperons
($i=\Lambda$, $\Sigma$; $k=n$, $p$, $\Lambda$, $\Sigma$).
%
%%%%%%%%%%%%%%%%%%%%%%%%%%% FIGURE 4  %%%%%%%%%%%%%%%%%%%%%%%%%%%%%
\begin{figure}[t]
\setlength{\unitlength}{1mm}
\leavevmode
\hskip  0mm
\includegraphics[width=150mm,bb=15  300  550  550,clip]{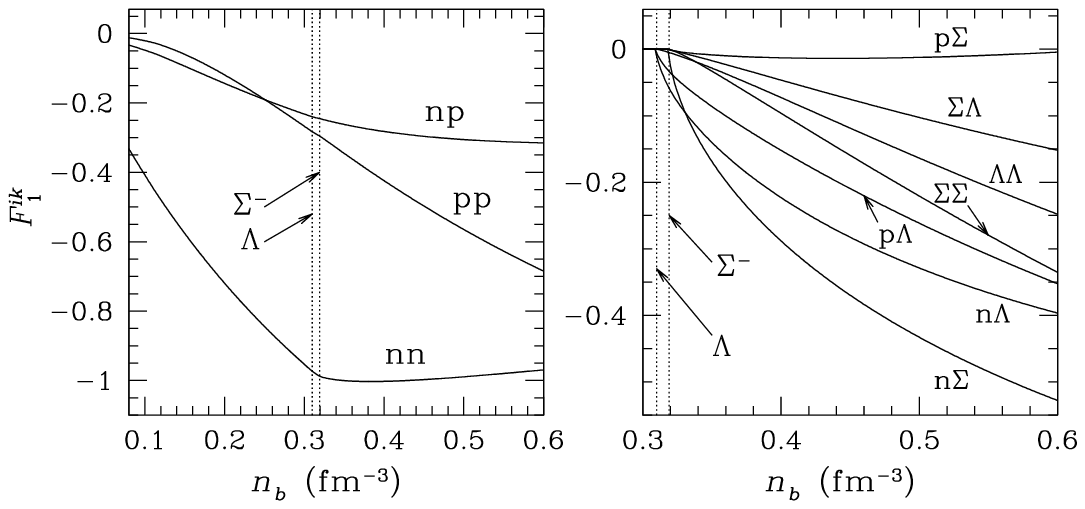}
\caption
{
The same as in Fig.\ 3 but for $F_1^{ik}$.
}
\label{4fig}   
\end{figure}
%%%%%%%%%%%%%%%%%%%%%%%%%%%%%%%%%%%%%%%%%%%%%%%%%%%%%%%%%%%%%%%%
%

We checked that the nucleon-hyperon matter is stable
down to baryon number density 
$n_b=0.34 n_0=0.055$ fm$^{-3}$ 
where the instability occurs 
(there are no hyperons and muons at such $n_b$).
Mathematically, the occurence of instability 
means that the inequality (\ref{usl2})
is not satisfied
at $n_b<0.34 n_0=0.055$ fm$^{-3}$.
Thus, the matter is unstable with respect 
to long-wavelength density fluctuations.
All other criteria, 
which are necessary for positive definiteness
of the quadratic forms (\ref{energy3}) and (\ref{dF1}), 
are obeyed.

This type of instability
is related to the crust-core phase transition
and is carefully analyzed 
in the neutron-star literature 
(see, e.g., \cite{bbp71, prl95, cyd97, 
dmc08, dpsbc08}).
Since we study the stability of matter 
only in the extreme long-wavelength limit
and under condition of microscopic charge neutrality,
our result for the baryon number density 
of the crust-core interface
is just the lower bound
for the real value.
Precise calculations would give a slightly 
higher value. 
For example, using extended Thomas-Fermi approach,
Cheng et al.\ \cite{cyd97} found 
the crust-core boundary 
at $(0.058-0.073)$ fm$^{-3}$,
depending on the choice of 
the $\sigma$-$\omega$-$\rho$ model parameters.
%
%$n_b \approx 0.5 n_0$.
%
%The chosen equation of state 
%leads to an instability of asymmetric nuclear matter
%at $n_b \leq 0.34 n_0 \approx 0.055$ fm$^{-3}$ 
%(there are no hyperons and muons at such $n_b$).
%Mathematically, this means that 
%inequality (\ref{usl2})
%in this case is not satisfied.
%It is important to note that all other criteria, 
%which are necessary for positive definiteness
%of the quadratic forms (\ref{energy3}) and (\ref{dF1}), 
%are obeyed.
%It is also worth noting that since we analyse the stability
%of matter only with respect to long-wavelength perturbations,
%our result for the occurence of instability
%is just the estimate from below for the real threshold.
%
%The region of $n_b$ where the uniform nucleon matter is unstable
%with respect to separation into a more dense 
%and a less dense component, 
%is shaded in Fig.\ 3.

%%%%%%%%%%%%%%%%%%%%%%%%%%%%%%%%%%%%%%%%%%%%%%%%%%%%%%%%%%%%%%%%%%%%%%%
\section{Summary}
\label{4}
%%%%%%%%%%%%%%%%%%%%%%%%%%%%%%%%%%%%%%%%%%%%%%%%%%%%%%%%%%%%%%%%%%%%%%%

In this paper we calculated
the relativistic entrainment matrix $Y_{ik}$ at zero temperature 
for nucleon-hyperon mixture
(see Eq.\ (\ref{Yik2})).
This matrix is a relativistic analogue of the 
entrainment matrix $\rho_{ik}$
(also termed the mass-density matrix 
or Andreev-Bashkin matrix)
and is related to $\rho_{ik}$ in the non-relativistic limit
by Eq.\ (\ref{nonrel_limit}).
The calculation is done in the frame 
of {\it relativistic} 
Landau Fermi-liquid theory \cite{bc76}, 
generalized to the case of mixtures.
We show that, similarly to $\rho_{ik}$ (see, e.g., \cite{bjk96,ch06}),
the matrix $Y_{ik}$ can be expressed 
through the Landau parameters $f_1^{ik}$ 
of nucleon-hyperon matter ($i, k=n$, $p$, $\Lambda$, $\Sigma$).
If the number of baryon species is more than four,
then the indices $i$ and $k$ in Eq.\ (\ref{Yik2}) 
should run over all these species.

The general results for $Y_{ik}$, following from
the relativistic Landau Fermi-liquid theory,
are illustrated with an example of 
the $\sigma$-$\omega$-$\rho$ 
mean-field model with scalar self-interactions.
Using this model we obtain the analytic expression
(\ref{Yik3}) for the matrix $Y_{ik}$.
Comparison of this expression with Eq.\ (\ref{Yik2}) 
allows to determine the Landau parameters $f_1^{ik}$
corresponding to the chosen mean-field model. 
Furthermore, we calculate the parameters $f_0^{ik}$ 
and find that all other (spin-averaged) Landau parameters
are equal zero, $f_{l}^{ik}=0$ at $l \geq 2$.
%Thus, the $\sigma$-$\omega$-$\rho$ model
%with scalar self-interactions is formulated
%in a language of relativistic Landau Fermi-liquid theory.

In addition, we formulate a number of stability criteria 
for beta-equilibrated nucleon-hyperon matter
%with respect to long-wavelength density fluctuations
(the positive definiteness of quadratic forms 
(\ref{energy3}) and (\ref{dF1})).
Employing the third equation of state of 
Glendenning \cite{glendenning85}, 
which is one of the versions of the $\sigma$-$\omega$-$\rho$ model
with scalar self-interactions,
%we found no instability
%of the nucleon-hyperon matter 
%of neutron stars down to the crust-core 
%interface.
%
we demonstrate that the nucleon-hyperon matter 
of neutron stars is stable
down to the crust-core interface. 
%
%at $n_b \approx 0.5 n_0$.
%the number density
%$n_b = 0.34 n_0 \approx 0.055$ fm$^{-3}$. 
%
%We demonstrate that 
%the third equation of state of Glendenning \cite{glendenning85}, 
%which is one of the versions of the $\sigma$-$\omega$-$\rho$ model
%with scalar self-interactions,
%leads to an instability of the ground state of nucleon matter
%at $n_b \leq 0.34 n_0 \approx 0.055$ fm$^{-3}$ 
%(no hyperons and muons at such $n_b$). 

Our results can be used to model the pulsations 
of cold massive neutron stars with 
superfluid nucleon-hyperon cores.
The generalization of these results 
to the case of finite temperatures
will be given in a subsequent publication.

%%%%%%%%%%%%%%%%%%%%%%%%%%%%%%%%%%%%%%%%%%%%%%%%%%%%%%%%%%%%%%%%%%%%%%%
\section*{Appendix A}
\label{App1}
%%%%%%%%%%%%%%%%%%%%%%%%%%%%%%%%%%%%%%%%%%%%%%%%%%%%%%%%%%%%%%%%%%%%%%%

Using Eqs. (\ref{jj})--(\ref{rho32}), 
one can express the particle current densities ${\pmb j}_i$ 
as functions of momenta ${\pmb Q}_k$, 
and thus derive the coefficients of 
relativistic entrainment matrix $Y_{ik}$ 
at zero temperature:
\begin{eqnarray}
Y_{ik}=\frac{n_i}{m_i^{\ast}}
\left[\delta_{ik}-\frac{g_{\omega i}}{A}
\frac{n_k}{m_k^{\ast}}\left(\frac{g_{\omega k}}{m_\omega^2}a_{22}
-\frac{g_{\rho k}I_{3k}}{m_\rho^2}a_{12}\right)
-\frac{g_{\rho i}I_{3i}}{A}\frac{n_k}{m_k^{\ast}}
\left(\frac{g_{\rho k}I_{3k}}{m_\rho^2}a_{11}
-\frac{g_{\omega k}}{m_\omega^2}a_{21}\right)\right].
\label{Yik3}
\end{eqnarray}
Here 
%$m_i^{\ast}=\sqrt{\pFa^2+(m_i-g_{\sigma i}\sigma)^2}$, 
$m_i^{\ast}$ is given by Eq.\ (\ref{effmassLandau})
while the coefficients $a_{11}, a_{12}, a_{21}, a_{22}$, and $A$ are given by
\begin{eqnarray}
a_{11}&=&1+\sum_i \frac{g_{\omega 
i}^2}{m_\omega^2}\frac{n_i}{m_i^{\ast}},\label{a11}\\
a_{12}&=&\sum_i \frac{g_{\omega i}g_{\rho 
i}I_{3i}}{m_\omega^2}\frac{n_i}{m_i^{\ast}},\label{a12}\\
a_{21}&=&\sum_i \frac{g_{\omega i}g_{\rho 
i}I_{3i}}{m_\rho^2}\frac{n_i}{m_i^{\ast}},\label{a21}\\
a_{22}&=&1+\sum_i \frac{g_{\rho i}^2 
I_{3i}^2}{m_\rho^2}\frac{n_i}{m_i^{\ast}},\label{a22}\\
A&=&a_{11}a_{22}-a_{12}a_{21}.
\end{eqnarray}
In formulae (\ref{a11})--(\ref{a22}) the summation 
is assumed over all baryon species.

%%%%%%%%%%%%%%%%%%%%%%%%%%%%%%%%%%%%%%%%%%%%%%%%%%%%%%%%%%
\section*{Acknowledgments}
%%%%%%%%%%%%%%%%%%%%%%%%%%%%%%%%%%%%%%%%%%%%%%%%%%%%%%%%%%

The authors are very grateful to
K.P. Levenfish and D.G. Yakovlev for allowing 
to use their code which calculates the third equation 
of state of Glendenning \cite{glendenning85}, 
to N. Chamel for reading the draft 
of the paper and valuable comments, 
and to anonymous referee for useful suggestions.
This research was supported in part
by RFBR (Grants 08-02-00837 and 05-02-22003),
by the Federal Agency for Science and Innovations
(Grant NSh 2600.2008.2), and 
by the Polish MNiSW (Grant N20300632/0450). 
Two of the authors (M.E.G. and E.M.K.) acknowledge
support from the Dynasty Foundation.
M.E.G. also acknowledges support 
from the RF Presidential Program 
(grant MK-1326.2008.2).

%%%%%%%%%%%%%%%%%%%%%%%%%%%%%%%%%%%%%%%%%%%%%%%%%%%%%%%%%%

\end{document}